\documentclass{JHEP}
\usepackage{epsfig}
\usepackage{amssymb}
\usepackage[active]{srcltx}

\newcommand{\bea}{\begin{eqnarray}}
\newcommand{\eea}{\end{eqnarray}}
\newcommand{\bean}{\begin{eqnarray*}}
\newcommand{\eean}{\end{eqnarray*}}

\def\O #1{\overline{#1}}

\def\ba{\begin{array}}
\def\ea{\end{array}}
\def\beq{\begin{equation}}
\def\eeq{\end{equation}}

\def\det{\mathop{\rm det}}

\def\a{{\alpha}}
\def\b{{\beta}}
\def\d{{\rm d}}

\def\th{{\theta}}
\def\bth{{\overline{\theta}}}

\def\half{{1 \over 2}}

\preprint{
{\tt hep-th/0307165}}
\title{${\cal N}=\half$ Wess-Zumino model is renormalizable
}
\author{Ruth Britto and Bo Feng\\
~~~~\\
Institute for Advanced Study \\
Einstein Drive, Princeton NJ 08540 \\
~~~~~~~~~~~~~~~~\\
\email{britto@ias.edu, fengb@ias.edu} }
\abstract{
The Wess-Zumino model on ${\cal N}=\half$ nonanticommutative superspace, which contains the dimension-6 term $F^3$,
 is shown to be renormalizable to all orders in perturbation theory, upon adding $F$ and $F^2$ terms to the original Lagrangian.  The renormalizability is possible, even with this higher-dimension operator, 
because the Lagrangian is not hermitian.  Such deformed field theories arise naturally in string theory with a graviphoton background.
}
\begin{document}
\section{Introduction}

It is commonly understood that for a four-dimensional quantum field theory to be renormalizable, every operator appearing in the Lagrangian must have a mass dimension less than or equal to 4.  An important assumption in this statement is that the action is real.  
However, it was recently observed \cite{newrefs} that
in string theory with a constant graviphoton background, the field theory of the string worldsheet suffers a deformation of its superspace coordinates $\th^\a$. These coordinates no longer anticommute, but satisfy a Clifford algebra: $C^{\a\b}=\{\th^\a,\th^\b\}$.  Supersymmetry is reduced by half.  It follows too that the action is no longer real.  In \cite{seiberg}, it was found that this deformation of superspace effectively adds new terms, of dimension greater than 4, to the Wess-Zumino and Yang-Mills Lagrangians.\footnote{This deformation of superspace was considered previously for example in \cite{non-theta}; recent work on ${\cal N}=\half$ supersymmetric field theories has appeared in \cite{recent,twoloop,bfr,ty,luninrey}.}
Initially it was expected that these deformed theories were unrenormalizable, but should simply be treated perturbatively.  In this letter we defy intuition by demonstrating the renormalizability of the ${\cal N}=\half$ Wess-Zumino model to all orders.  Our result generalizes the recent work \cite{twoloop}, which proved renormalizability up to two loops.

The classical Lagrangian is 
\bea  \label{Lag}
\int\d^4\th~\O\Phi\Phi + \int\d^2\th~({m \over 2}\Phi^2 + {g \over 3}\Phi^3)+\int\d^2\bth~({\O m \over 2}\O\Phi^2 + {\O g \over 3}\O\Phi^3)~~-~{g \over 3}\det(C)F^3,
\eea
where 
\bean
\Phi(y,\th) &=& A(y) + \sqrt{2}\th\psi(y) + \th\th F(y) \\
\O\Phi(\O y,\O\th)&=&\O A(\O y) + \sqrt{2}\O\th\O\psi(\O y) + \O\th\O\th\O F(\O y),
\eean
and $y$ and $\O y$ are chiral coordinates.
The $F^3$ term arising from the deformation of superspace might appear to render the theory unrenormalizable.  However, the recent paper \cite{twoloop}
showed that up to two loops it is in fact renormalizable, after adding by hand $F$ and $F^2$ terms to the Lagrangian.
Specifically, \cite{twoloop} proved the following assertions up to two loops:
\begin{itemize}
\item Divergences are at most logarithmic.
\item Divergent terms have at most one power of $\det(C)$.
\item Divergent terms are of the form $F^\ell \O G^k$, where $\O G \equiv 
\O m \O A + \O g \O A^2$.
\item Contractions with $\O G$ are equivalent to contractions with $F$, and thus the counterterms from $F,F^2,F^3$ suffice to renormalize the theory.
\end{itemize}
In this work we prove the same assertions to all orders. We work in terms of component fields, making use of the two $U(1)$ (pseudo)symmetries of the theory (see \cite{bfr} and the Appendix), as well as Feynman diagram combinatorics, to constrain the form of divergent terms in the effective action. 

The reasoning of this kind of renormalizability is discussed in detail in \cite{luninrey}.
The essential point is that the higher-dimensional term (like $F^3$)
acts as a source vertex with  no corresponding sink (like 
$\O F^3$ or $A^3$), so these vertices  do not show up in arbitrarily 
high order diagrams. This point can be seen clearly in our argument below;
for example, the $F^3$ term can appear only  once in any divergent
diagram.

\section{At most log divergence and one power of $\det(C)$}\label{sec-c}

In this section we use symmetry arguments to show that the effective action has at most logarithmic divergences, and that the divergent terms have at most one power of $\det(C)$.

Suppose a general divergent operator appears in the effective action with a coefficient $\lambda$:
\bean
\Gamma_{\cal O} = \lambda {\cal O},
\eean
and that $\lambda$ has mass dimension $d$ and charges $q_R = R, q_{\Phi}=S$ under the two global U(1) (pseudo)symmetries, whose action is described in the Appendix.
By dimensional analysis 
 and charge considerations, and with an ultraviolet momentum cutoff, we will have
\bea\label{lambda}
\lambda\sim \Lambda^d g^{x + 4 z-R} \O g^{x}
\left( {m \over \Lambda} \right)^y
\left( {\O m \over \Lambda} \right)^{ y + 6 z +{S-3R\over 2}} 
\left( \det(C) \Lambda^2\right)^z 
\eea
from the loop calculation, where $x,y,z$ are nonnegative integers; the nonegative powers of $m,\O m,g,\O g$ and $\det(C)$  come from the propagators and vertices.

The most general operator  ${\cal O}$ is given by
\bean
{\cal O} \sim \partial^p A^{\alpha_1} \O A^{\O \alpha_1} F^{\alpha_2}
\O  F^{\O \alpha_2} \psi^{\alpha_3} \O \psi^{\O \alpha_3}.
\eean
The differential operators are understood to act on the others in all possible combinations; for our purposes it is enough to count the overall dimension.
Because the term $\Gamma_{\cal O}$ must have mass dimension 4 and zero charge, we must have 
\bean
d & = & 4- p - ( \alpha_1+ \O \alpha_1+ 2 \alpha_2+ 2\O \alpha_2
+{3\over 2} \alpha_3+{3\over 2}\O \alpha_3)\\
R & = & -\alpha_1+ \O \alpha_1+  \alpha_2 -\O \alpha_2 \\
S & = &  -\alpha_1+ \O \alpha_1-\alpha_2+\O \alpha_2
- \alpha_3+\O \alpha_3
\eean
Thus, the overall power of $\Lambda$ in $\Gamma_{\cal O}$ is
\bea
P & = & 4-4z-2y - p - 2\alpha_1- 4 \O \alpha_2-\alpha_3-2 \O \alpha_3.
\eea
The new operator is superficially divergent iff $P\geq 0$.  Note that this restricts $z$, which is the power of $\det(C)$, to be 0 or 1.  (1) If $z=0$, it is the ordinary Wess-Zumino case with only the familiar 
wave function renormalization; we will not discuss it here. 
(2) If 
$z=1$, then 
$y=p=\alpha_1=\O \alpha_2=\alpha_3=\O \alpha_3=0$, and 
there is at most a logarithmic divergence.  The possibly divergent operators take the form $\O A^{\O \alpha_1} F^{\alpha_2}$.

\section{Finitely many divergent operators}\label{sec-finite}

Now we show that there are only finitely many divergent operators, namely 
\bea
\O A^{\O \a_1} F^{\a_2}, \qquad {\rm for}~~\O \a_1 + 2 \a_2 \leq 6~~{\rm and}~~\a_2 \geq 1. 
\eea
Let us examine the Feynman diagrams that generate a given operator.
We introduce the following nonnegative integers to count the number of 
each type of vertex and propagator that appears in the diagram:
\bean
a_0:~~F^3,~~~~a_1:~~\O A^2 \O F,~~~~b_1:~~\O A\O \psi \O \psi,~~~~
a_2:~~A^2 F,~~~~b_2:~~A\psi \psi,
\eean
\bean
& & c_1:~~A\O A,~~~~c_2:~~F \O F,~~~c_3:~~AF,~~~~c_4:~~\O A\O F,\\
& & d_1:~~\psi \psi,~~~~d_2:~~\O \psi \O \psi,~~~~d_3:~~\psi\O \psi.~~~~
\eean
Now, by matching powers of the coupling constants, we can specify the exponents in the previous section: $(x,y,z) = 
(a_1 + b_1,c_4+d_2,a_0)$.  Moreover, restricting to possibly divergent operators, we apply the main result of the previous section to set $c_4=d_2=0$ and $a_0=1$.
By matching the appearances of each field among vertices, contractions and external lines, we can derive six conditions on these integers.
One consequence of these conditions is that $d_1  =  6-\O \a_1 -2 \a_2 -c_3$.  Because $d_1$ is nonnegative, it follows that $\O \a_1 + 2 \a_2 \leq 6$, which is the desired result.  The additional condition $\a_2 \geq 1$ follows from the nonrenormalization of the antiholomorphic superpotential \cite{bfr,ty}.
 
\section{Repackaging in terms of $\O G$}

The arguments of the previous two sections have proved the renormalizability of the ${\cal N}=\half$ supersymmetric Wess-Zumino model.
However, the observation made in \cite{twoloop} up to two loops is a stronger result:  
{\it  divergent terms can be packaged in terms of $F$ and 
$\O G \equiv \O m \O A + \O g \O A^2$}. 
We now show that this result applies generally.
 The divergent terms generated are 
\bean
F, ~~~F^2, ~~~F^3, ~~~F \O G, ~~~F^2 \O G,~~~F \O G^2.
\eean
We then invoke the beautiful argument of \cite{twoloop}, that a contraction of any field with $\O G$ is equivalent to its contraction with $F$.\footnote{Being able to trade $\O G$ exactly for $F$ is also a consequence of the nonrenormalization of the antiholomorphic superpotential that was shown in \cite{bfr,ty}.}  
We conclude that the counterterms for $F, F^2$ and $F^3$ suffice to renormalize the theory.

To begin, we simplify our task by neglecting diagrams with fermion loops. This is allowed, because there will be a corresponding bosonic loop that cancels the contribution at least partially.  Since the leading order is already log divergent, partial cancellation gives at most a finite total contribution.  So we are free to set $b_i=d_j=0$.

We will classify our divergent 1PI diagrams by powers of $\O A$ (as external lines) relative to a base diagram with no powers of $\O A$.
Consider the following operations on diagrams: 
\begin{itemize}
\item Insertion:  Where there is a propagator $AF$, the vertex $\O A^2 \O F$ can be inserted, so that there are now two propagators, $A \O A$ and $F \O F$, and an additional external line $\O A$.  It is easy to see that
the divergence from the momentum integration is not modified.
The replacement can be sketched as $\O m \rightarrow \O g \O A$.

\item Deletion:  Where there is an external line $\O A$, it must lead to a vertex $\O A^2 \O F$, since there are no fermion loops.  Because $c_4=0$, there are no propagators $\O A\O F$.  Thus this vertex must be contracted with $A$ and $F$, via the propagators $A \O A$ and $F \O F$.  The vertex can be cut out of the diagram, and the two propagators replaced by the single propagator $AF$.  Again, the divergence is not modified.
\end{itemize}
These two operations are inverse.  By applying deletion repeatedly, every diagram can be reduced to a `base diagram' with only $F$ on the external lines. To see these two operations more clearly, Figure \ref{f:cutpaste}
will be helpful.

\EPSFIGURE[ht]{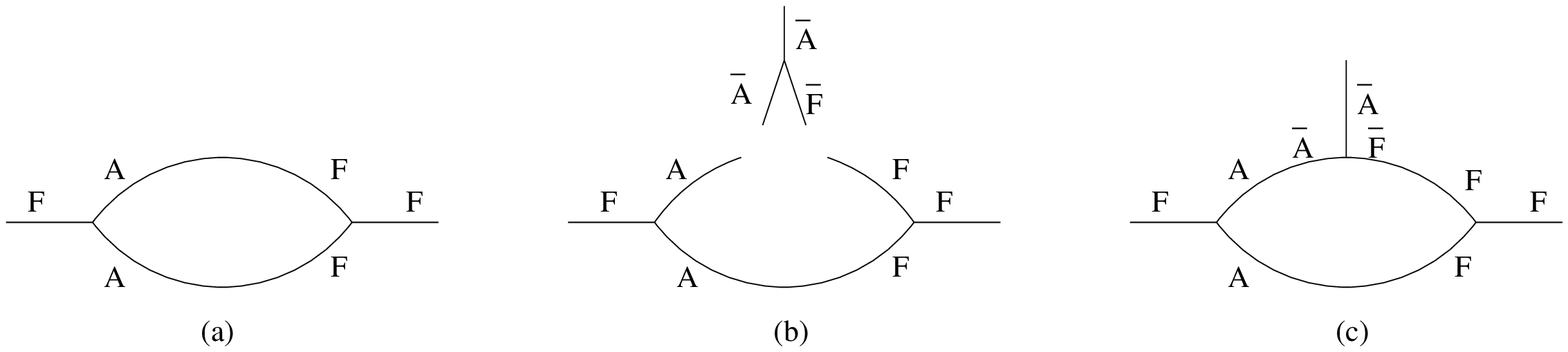,width=12cm}
{ In this figure, we demonstrate how $\O A F^2$ is related to
$ F^2$ by the insertion and deletion operations.
\label{f:cutpaste}
}

We apply insertion to build up all possible diagrams with a given base diagram for $F^\ell$.  
Let $q$ be the number of $AF$ propagators in the base diagram.  From the results in the previous section, we know in fact that $q=6-2\ell$. Note in particular that $q$ is even and $q \leq 4$.  

From this base, there are ${q \choose k} 2^k$ ways to insert $k$ external lines  $\O A$---namely ${q \choose k}$ ways to choose the propagators for insertion, and two choices for the external line from each new vertex $\O A^2 \O F$.  Moreover, each insertion yields a diagram with a relative factor ${\O g \over \O m}.$

Relative to the base diagram including its symmetry factor $S$, the sum of all the contributions with $k \geq 1$ is
\bea
S F^\ell \sum_{k=1}^q {q \choose k}2^k ({\O g \over \O m})^k \O A^k,
\eea
which is simply 
\bean
S F^\ell \left[(1 + 2 {\O g \over \O m}\O A)^q -1 \right].
\eean
Since $q=2(3-\ell)$, 
\bean
(1 + 2 {\O g \over \O m}\O A)^q -1
= (1 + 4 {\O g \over \O m^2}\O G)^{(3-\ell)} -1.
\eean
Thus the sum over contributions from diagrams with $k\geq 1$, with a given base diagram, is indeed a polynomial in $\O G$, namely
\bea
S F^\ell \sum_{k=1}^{3-l} {3-l \choose k}4^k ({\O g \over \O m^2})^k \O G^k.
\eea


\section{No further divergences from $F$ and $F^2$ vertices}

To renormalize the theory, we must add the terms $F$ and $F^2$  by hand to the tree-level Lagrangian. We choose to parametrize these terms as
\bea
\lambda_1 g^3\O m^4 \det(C) F + \lambda_2 g^2 \O m^2 
 \det(C) F^2
\eea
The advantage of this parametrization is that both 
$\lambda_1, \lambda_2$ are dimensionless and charge neutral under
both $U(1)_R$ and $U(1)_\Phi$.

It is easy to see that in this general situation, the symmetry
argument given in section two still applies. For example,
we only need to add $\lambda_1^{w_1}\lambda_2^{w_2}$ to 
equation (\ref{lambda}), which does not affect the power counting.
We still have following conclusions. (1) A log divergent term has at most one $\det(C)$ insertion.
This
means a divergent diagram contains at most one of any of the three
vertices $F, F^2, F^3$. (2)  The only possible divergent operators, with one power of $\det(C)$, are 
still 
\bea
\O A^{\O \alpha_1} F^{\alpha_2}, \qquad {\rm for}~~
\O \a_1 + 2 \a_2 \leq 2 \ell,~~~~\ell=1,2,3
\eea
where the vertex $F^{\ell}$ is the one appearing in the diagram. 


The crucial point is that with two new terms $F, F^2$ in the
tree level Lagrangian, the insertion and deletion operations
are not modified at all. The repackaging of all these divergent
terms into $\O G$ will go through unmodified. 
Thus we conclude that the renormalizability proved in the previous
sections continues to hold with the  terms $F, F^2$ in the
tree level Lagrangian.

\section*{Acknowledgements}

We thank O. Lunin, S. J. Rey, and N. Seiberg for helpful discussions.  This work was supported by the NSF grant PHY-0070928.
%
%


\appendix

\section*{Appendix:  The two $U(1)$ (pseudo)symmetries}

For the ${\cal N}=\half$ Wess-Zumino model, we can identify two global $U(1)$
(pseudo)symmetries
by treating all coupling parameters, including the nonanticommutativity parameter $C^{\a\b}$, as the lowest components of (anti)chiral
superfields. They are $U(1)_\Phi$ flavor symmetry and $U(1)_R$
R-symmetry. Charge assignment is given as follows \cite{bfr}.

\begin{center}
\begin{tabular}{|c|c|c|c|c|c|c|c|} \hline
 & dim & $U(1)_R$ & $U(1)_\Phi$ &  & dim & $U(1)_R$ & $U(1)_\Phi$\\\hline
$\th$ & -1/2 & 1 & 0 & $\O \th$ &  -1/2 & -1 & 0 \\ \hline
$d\th$ & 1/2 & -1 & 0 & $d\O \th$ &  1/2 & 1 & 0 \\\hline
$A$ & 1 & 1 & 1 & $\O A$ & 1 & -1 & -1\\\hline
$\psi$ & 3/2 & 0 & 1 & $\O \psi$ & 3/2 & 0 & -1 \\\hline
$F$ & 2 & -1 & 1 & $\O F$ & 2 & 1 & -1 \\\hline
$g$ & 0 & -1 & -3 & $\O g$ & 0 & 1 & 3 \\\hline
$m$ & 1 & 0 & -2 & $\O m$ & 1 & 0 & 2 \\\hline
$C^{\alpha \beta}$ & -1 & 2 & 0 & $|C|$ & -2 & 4 & 0  \\\hline \hline
$\Phi$ & 1 & 1 & 1 & $\O \Phi$ & 1 & -1 & -1 \\\hline
\end{tabular}
\end{center}


\bibliographystyle{JHEP}

\end{document}